\documentclass{article}[12pt,a4,pre,preprint]
\usepackage{amsmath}
\usepackage{latexsym}
\usepackage{float}
\usepackage{amssymb}
\usepackage{graphicx}
\usepackage{cite}
\usepackage{epsfig}
\usepackage{graphics}

\begin{document}

\title{Molecular Dynamics simulation of evaporation processes of fluid bridges confined in slit-like pore}

\author{%
        Katarzyna Bucior, $^{\text{}}\;$%
   Leonid Yelash, $$%
         and Kurt Binder $^{\text{}}$
                          \\[\baselineskip]%
                   $^{\text{}}$ \textit{ Institut f\"{u}r Physik, Johannes Gutenberg-Universit\"{a}t} \\
                   \textit {D-55099 Mainz, Staudinger Weg 7, Germany}\\%
                   \date{}
}

 \maketitle

\begin{abstract}
A simple fluid, described by point-like particles interacting via
the Lennard-Jones potential, is considered under confinement in a
slit geometry between two walls at distance $L_z$ apart for
densities inside the vapor-liquid coexistence curve. Equilibrium
then requires the coexistence of a liquid ``bridge'' between the two
walls, and vapor in the remaining pore volume. We study this
equilibrium for several choices of the wall-fluid interaction
(corresponding to the full range from complete wetting to complete
drying, for a macroscopically thick film), and consider also the
kinetics of state changes in such a system. In particular, we
study how this equilibrium is established by diffusion processes,
when a liquid is inserted into an initially empty capillary
(partial or complete evaporation into vacuum), or when the volume
available for the vapor phase increases. We compare the diffusion
constants describing the rates of these processes in such
inhomogeneous systems to the diffusion constants in the
corresponding bulk liquid and vapor phases.
\end{abstract}

\section{Introduction}

Fluids confined into pores with diameters on the micrometer scale
down to the nanometer scale are important in a variety of
contexts: they control the properties of wet granular matter
\cite{1}, they play a role for oil recovery from porous rocks or
compactified sands \cite{2}, separation processes in zeolites
\cite{3}, drying processes of food, wood, or other porous solids
\cite{4}, nanofluidic devices using fluids in carbon nanotubes
\cite{5}, dip-pen nanolithography \cite{6}, nanolubrication
\cite{7}, fluid transport in living organisms \cite{8}, etc.
\cite{9,10,11,12}. Many of these applications involve
nonequilibrium processes such as the flow of fluids in confined
geometry, imbibition of fluids into pores (e.g.
\cite{13,14,15,16,17}), surface-directed spinodal decomposition
(e.g. \cite{18,19,20,21,22}) if one considers binary fluid
mixtures, or evaporation processes of fluids \cite{4}. While the
evaporation process of bulk liquids (across a flat interface)
\cite{23,24,25,26,27,28,29} and of small droplets
\cite{30,31,32,33} has been studied extensively, liquid-vapor
transitions under confinement have been mostly studied emphasizing
equilibrium aspects only \cite{9,12,34,35,36,37,38,39,40,41}. Note
that we do not discuss here the inverse process (nucleation of
fluid droplets from the vapor in confined systems, see e.g.
\cite{42,43}), and we consider neither the structure (and possible
rupture) of non-volatile confined liquid bridges \cite{44,45}, nor
liquid-vapor systems confined by patterned surfaces (see e.g.
\cite{44,45,46,47}.

In the present paper we wish to contribute to the understanding of
the kinetics of nonequilibrium processes of confined fluids
considering the partial (or complete) evaporation of liquid
bridges, resulting from changes of external conditions, using
Molecular Dynamics methods \cite{48,49,50} to simulate a simple
Lennard-Jones fluid confined between parallel walls. We are
particularly interested in elucidating the consequences of the
inhomogeneity of the structure of the liquid bridges (depending on
the drying/wetting boundary conditions at the 
confining walls \cite{9,51,52,53,54,55,56}, there is also an
inhomogeneity of the system in the $z$-direction perpendicular to
the walls) on the evaporation processes.

In the following section, we shall describe the model that is
simulated and give a few comments on the simulation method, and
the quantities, that are computed. In Sec. III, we describe the
static structure of the liquid bridges (as well as the regions of
the slit pore where the vapor phase dominates), for various
choices of the strength of the interaction between the fluid
particles and the walls, relating the observed interfacial
behavior to theoretical concepts about wetting. In Sec IV, we
present some discussion on the transport behavior of pure
coexisting vapor and liquid phases, under confinement in the slit
pores. Sec. V then considers the relaxation of the system after
suitable parameter changes that lead to partial (or even complete)
evaporation of the liquid bridges present in the slit pores. The
kinetic evolution towards the new (inhomogeneous!) equilibrium is
documented in detail. Finally, Sec. VI contains our conclusions,
and gives an outlook on future work.

\section{Model and Simulation Method}

In this study we are not addressing a particular material but
rather wish to gain insight into the generic behavior of simple
fluids (such as CH$_4$, CO$_2$, etc.) under confinement. Thus, we
describe the fluid particles as point-like, interacting via
truncated and shifted Lennard-Jones potential,

\begin{eqnarray} \label{eq1}
&& U(r)=U_{LJ} (r) - U_{LJ} (r_{\rm cut}), \nonumber\\
&&U_{LJ}= 4 \varepsilon [(\sigma/r)^{12} - (\sigma/r)^6 ] \, ,
\end{eqnarray}

with $r_{\rm cut} = 2 r_{\rm min}$, $r_{\rm min}=2^{1/6}\sigma$.
Note that $U_{LJ} (r_{\rm cut})$ is chosen such that $U(r)$ is
everywhere continuous, with $U(r \geq r_{\rm cut})=0$. Mognetti et
al. \cite{57} have shown that this model can describe fairly well
the coexistence curves, vapor pressure, and interfacial tension of
molecules like CH$_4$ or even C$_3$H$_8$ over a temperature regime
from about 0.7 $T_c$ to $T_c$, when $\varepsilon$ and $\sigma$ in
Eq.~(\ref{eq1}) are adjusted such that the critical temperature
$T_c$ and critical density $\rho_c$ are correctly reproduced by
the model ($\rho_c$, $T_c$ can be accurately estimated from the
model by careful finite size scaling analyses of Monte Carlo
simulations of the model in the grand-canonical ensemble, as
discussed elsewhere \cite{57,58}). Even for CO$_2$ this model
yields a fair description \cite{58}, although better accuracy can
be obtained for this molecule if the quadrupolar interaction is
included \cite{57}, but this is out of consideration here.

For a fluid confined in a slit pore geometry, it is also necessary
to specify the boundary conditions created by the planar walls
confining the thin film. Following Ref. \cite{22}, we choose an
atomistic description of these walls, setting particles on a
regular (and rigid) triangular lattice of lattice spacing $\sigma$, 
in the ($x,y$) plane at $z=1$ and at $z=L_z-1$ (note that
henceforth lengths are measured in units of $\sigma$ throughout).
The interaction between wall particles and fluid particles is
chosen also of the form of Eq.~(\ref{eq1}), but the energy
parameter $\varepsilon$ is replaced by $\varepsilon_w=\zeta \varepsilon$
with $\zeta$ being varied from $\zeta=0.1$ to $\zeta=0.9$. An
additional simulation was made where we made the interaction
between the wall and fluid particles purely repulsive: in this
case, the wall particles were placed in the planes $z=0.5$ and
$z=L_z - 0.5$, respectively, and $r_{\rm cut} =r_{\rm min}$ was
chosen in Eq.~(\ref{eq1}), so that $U_{LJ} (r_{\rm cut})=0$ then.
In this case, $\varepsilon_w=\varepsilon$ was used. Henceforth, we
shall also measure temperature in LJ units (i.e., $\varepsilon
\equiv 1$, $k_B \equiv 1)$. Typical linear dimensions in $x$ and
$y$ directions parallel to the walls were at least $L_x=30$,
$L_y=15$, choosing periodic boundary conditions in these
directions, and also $L_z=15$ or $L_z=16$ was chosen: the reason
for choosing $L_x=2L_y$ is that then the vapor-liquid interfaces
that form in the slit pore for suitable total densities will be
oriented in the $yz$-plane, perpendicular to the $x$-direction; to
minimize the interfacial free energy cost. No inhomogeneity in
$y$-direction is then expected, and hence averages of density
profiles along the $y$-direction can be taken. When liquid bridges
occur, the density distribution must then depend on both the $x$
and the $z$ coordinates.

Of course, the choice of the linear dimensions in a computer
simulation is a subtle matter, due to finite size effects
\cite{50,59}. While the finite size effects associated with the
two walls defining the finite width of the slit pore are the
physically significant effect that we wish to study, finite size
effects due to too small values of $L_x$ and $L_y$ are simulation
artefacts, since the real systems (if they are finite of
nanoscopic size in $x$ and $y$ direction as well) would have other
boundary effects rather than periodic boundary conditions. Choice
of the latter makes sense to simulate a system that is truly
macroscopic in $x$ and $y$ directions, of course. While
homogeneous systems (containing a single phase) typically approach
the thermodynamic limit rapidly, at least for temperatures or
densities outside of the critical region \cite{59}, this is not
always the case for systems exhibiting phase coexistence within a
finite simulation box \cite{38,59,60,61}. In particular, when we have
a liquid slab with two interfaces (perpendicular to the
$x$-direction) in our system, we expect that both interfaces have
a finite thickness: the standard description postulates an
``intrinsic thickness'' $\ell = 2 \xi$, where $\xi$ is the
correlation length of density fluctuations in the liquid,
broadened by capillary waves \cite{62}. While $\xi$ diverges near
the critical point of the fluid, $\xi \approx \sigma$ if we are
far below criticality. However, since the broadening caused by
capillary waves increases logarithmically with $L_y$, an
interfacial thickness of $2 \sigma$ to $4 \sigma$ must be typically
expected. However, the two parallel interfaces separating the
liquid bridge from the surrounding vapor in the simulation box
need to be at a distance $L\gg \ell$, in order that interactions
between these interfaces are negligible. Because of the periodic
boundary condition in $x$ direction, $L_x$ must exceed $L$ by a
factor of 2 or 3 as well. In order to test for possible finite
size effects, we have done part of our simulations with larger
linear dimensions, up to $L_x=90$.

All simulations were carried out using Molecular Dynamics methods
in the framework of the NVT ensemble. Applying the ESPResSO
package \cite{63}, the Newton equations of motion are integrated
via the Velocity Verlet algorithm \cite{48,49,50}, using a time
step $\delta t=0.002 \tau$ where the MD time unit $\tau$ is
defined as $\tau=\sigma (m/\varepsilon)^{1/2}$, where the mass
$m=1$ is chosen for the molecules. Temperature was controlled by
using the Langevin thermostat and $T^*=0.9366$ was chosen throughout.

In order to produce initial states containing liquid bridges
surrounded by vapor, we first equilibrated dense fluids (average
density $\rho\approx 0.5813$) in a small box with $L_x=30$,
$L_y=20$, $L_z=16$ and 9000 particles. 
The density of the fluid is calculated as $\rho=N\sigma^3/V$, where $N$- number of particles
 and $V=L_xL_y(L_z-2)$ for attractive walls, and $V=L_xL_y(L_z-1)$ for repulsive walls.
After an equilibration run
extending over at least $100 \tau$ we put this liquid in the center of a box
with linear dimension $L_x=60$, $L_y=20$, $L_z=16$, and simulate
this system over a time interval of 20000 $\tau$. The initial
stages of such a run serve as a simulation of liquid evaporation
into vacuum. For time $t > 5000 \tau$, the liquid bridge is
essentially well equilibrated with respect to the coexisting vapor, 
and then averages of the density
profiles $\rho (x,z)$ are taken.

\section{Static structure of the liquid bridges}

In order to obtain an  overview of the behavior,
Figs.~\ref{fig1},~\ref{fig2} present contour plots of the density
distribution $\rho(x,z)$ for systems of sizes $L_x=60$, $L_y=20$
and $L_z=16$ and $L_x=90$ $L_y=20$ and $L_z=16$, respectively. In
both cases the color coding goes from dark blue, corresponding to
$\rho=0$, to red, corresponding to $\rho=0.65  $, as indicated by
the bar on the top of each figure. The individual
pictures illustrate the variation with $\varepsilon_w$, while
$N$=9000 particles were used throughout. Thus, the average density
in Fig.~\ref{fig1} is $\bar{\rho} \approx0.291$ (when we use the
maximum available volume $60 \times 20 \times 14$, remembering
that the wall particles are fixed at $z=1$ and at $z=L_z-1=15$,
respectively), while in Fig.~\ref{fig2} it is only $\bar{\rho}
\approx 0.194$ due to the larger $L_x=90$. 
Note that at the chosen temperature $T^*=0.9366$ the
coexisting vapor and liquid densities are $\rho_v=0.109$,
$\rho_\ell=0.565$; thus, if the density distribution would be
homogeneous in the $z$-direction across the slit pore, we could
simply expect a two-phase equilibrium with a volume fraction
$X=(\bar{\rho} - \rho_v) / (\rho_\ell - \rho_v)$ of the liquid
phase, and the vapor-liquid interfaces would simply show up as
straight lines in the $z$-direction in
Figs.~\ref{fig1},~\ref{fig2}. However, due to the particle-wall
interactions, a nontrivial inhomogeneity of the density
distribution in the $z$-direction results, which readily shows up
in Figs.~\ref{fig1},~\ref{fig2}.

For small $\varepsilon_w$ one sees in Figs.~\ref{fig1},~\ref{fig2}
clear evidence of drying behavior: rather than observing a liquid
bridge connecting the confining walls, there occurs in the center of the slit pore  a free
standing elliptic liquid cylinder periodic in y direction, separated by
the vapor phase that has intruded in between the liquid and the
wall. 
%Considering the fact that the scale in the $x$-direction is
%compressed by about a factor 3.2 in Fig.~\ref{fig1} and 4.7 in Fig.~\ref{fig2}
% relative to the scale in the
%$z$-direction, 
It is obvious that the vapor-liquid interfaces in
the $z$-direction actually are distinctly narrower than along the
$x$-direction. This effect in fact is expected, of course, since
confinement has a strong constraining effect on interfacial
fluctuations in the $z$-direction, while the finite size $L_x$ in
$x$-direction has much less effect to constrain the interfaces.

While in the case of $L_x=60$ the linear dimension of the liquid
slab in the $x$-direction is large enough, so that in the center
of the slab the dependence of any physical properties on the
$x$-coordinate is negligible small, for $\varepsilon_w=0.1$ to
$\varepsilon_w=0.4$, this clearly is not true for $L_x=90:$ as
$\varepsilon_w$ increases the linear dimension of the liquid slab
along the $x$-axis gets smaller and smaller (since a larger
fraction of particles now stays in the gas phase, and for larger
$\varepsilon_w$ more and more particles get adsorbed at the
walls). The elliptical shape of the contours of constant density
in Fig.~\ref{fig2} clearly imply that the two
interfaces separating vapor from liquid and then liquid from vapor
along the $x$-direction interact with each other. We do expect
that such thin liquid slabs with interacting interfaces can
evaporate much easier than thick slabs, where these interfaces are
separated by a thick region of bulk liquid from each other.

For $\varepsilon_w=0.5$ the vapor no longer can intrude in between
the liquid slab and the walls, rather the liquid-vapor interface
is ``cut off'' by the walls. In the region from
$\varepsilon_w=0.61$ to $\varepsilon_w=0.65$ (the last one not shown), 
the vapor-liquid
interface seems to run almost straight along the $z$-direction,
implying (in the macroscopic limit of very thick slabs) a contact
angle of $\theta=90^o$, while for $\varepsilon_w \leq 0.5$ one
clearly can speak about a contact angle exceeding 90$^o$ (as long
as $\theta < 180^o$ one speaks about ``incomplete drying'', while
a thick vapor region in between the walls and the fluid slab
should correspond to $\theta=180^o$, complete drying). For
$\varepsilon_w\geq 0.7$ the shape of the liquid slabs, connecting
the confining surfaces in Fig.~\ref{fig1} suggest contact angles
$\theta < 90^o$, corresponding to incomplete wetting conditions.
However, we emphasize that for nanoscopically thin films the
concept of a contact angle is somewhat ill-defined, since it
requires that over distances much larger than the interface
thickness its curvature is negligible. This condition clearly is
not satisfied here. This lack of a precise definition of the
contact angle in nanoscopically thin slits corresponds to the fact
that wetting transitions in such geometries show finite-size
rounding \cite{64}.

One can also recognize from Fig.~\ref{fig1} for $\varepsilon \geq
0.61$ a pronounced layering effect near the walls. This layering
effect obscures the ``contact region'' where the interface meets
the wall. For $\varepsilon=0.9$, the walls are coated already with
precursors of wetting layers, indicative of the vicinity of the
rounded wetting transition. Note, however, that for
$\varepsilon_w=0.9$ the total particle number in the system does
not suffice to allow the formation of a well-defined liquid bridge
reaching bulk liquid density in the center of the system. For the
larger system $(L_x=90)$, the liquid bridge already has
disappeared somewhere in between $\varepsilon_w=0.65$ and $0.70$,
since there is enough space for all the particles to either stay
in the vapor or get adsorbed in the precursors of the wetting
layers near the walls.

At this point, we mention that in our evaluation of simulation data we always have
fixed the center of mass of all the particles right in the center
of the slit pore (i.e., at $x=L_x/2)$. Due to the periodic
boundary condition in $x$-direction, translational invariance in
$x$-direction is implied of course. Thus, fixing the center of
mass is a convenient precaution against the diffusion of the
liquid along the $x$-axis as a whole, which would smear our the
density inhomogeneity on average, of course; such an undesirable
effect could obscure corresponding experimental observations,
where one cannot control the position of the liquid slab as
easily.

Fig.~\ref{fig3} shows two selected profiles
$\rho(x_1,z)$ and $\rho(x_2,z)$ in more detail, choosing $x_1$
such that a cut through the center of the liquid slab is
performed, while $x_2$ is chosen to monitor the density profile
through the slit pore in the vapor region, far away from the
liquid slab. 
In each part of Fig.~\ref{fig3} there are presented plots of density for 
2 system sizes: $L_x=60, L_y=20, L_z=16$ (solid lines) and 
 $L_x=90, L_y=20, L_z=16$ (lines with circles).
Density profiles for both box sizes $L_x$ overlap for 
values of parameter $\varepsilon_w<0.65$.
One can see that in the region of incomplete drying
($\varepsilon_w=0.1$) the liquid density decreases almost linearly
with $z$ over a significant region of $z$ when one approaches
either wall. 
For $\varepsilon_w=0.3$ there is already evidence of
a layering effect in the density profile. While in the
vapor phase the density is higher in the layer adjacent to the wall than in
the center of the slit pore, in the liquid phase the behavior is
different - the density in this region is significantly higher then 
in the layer close to the wall.
However for $\varepsilon_w=0.5$ the density in the center of
the slit pore is already lower than in the layers adjacent to
the walls. 
%For $\varepsilon_w =0.5$, the density peaks
%corresponding to the first two layers are of exactly the same
%height as the density in the bulk of the liquid slab, while
For
$\varepsilon_w \geq 0.7$ the layering effect is even more
pronounced, leading to a density  
that is higher even in the second layer
adjacent to the wall than in the center of the slit pore.

In order to characterize the behavior more precisely in the region of
those values of $\varepsilon_w$ where the interfaces between vapor
and liquid are approximately planar, Fig.~\ref{fig5} presents
magnified plots of the profiles $\rho(x_1,z)$ and $\rho(x_2,z)$
vs. $z$ where 5 values of $\varepsilon_w$ from $\varepsilon_w=0.4$
to $\varepsilon_w=0.65$ are shown. This situation where the
contact angle $\theta=90^o$ is the ``transition'' from incomplete
drying to incomplete wetting \cite{51,52,53,54,55,56}. We
emphasize, however, that even in the thermodynamic limit ($L_z
\rightarrow \infty$) this ``transition'' is a completely smooth
change, unlike the wetting transition (where $\theta \rightarrow
0)$ or the drying transition (where $\theta \rightarrow 180^o$),
which become sharp thermodynamic transitions (singularities of the
surface excess free energies associated with the walls
\cite{51,52,53,54,55,56}) in this limit. We also note that the
density profile in the vapor phase becomes almost horizontal
already for $\varepsilon_w=0.4$, while the density profile in the
center of the liquid slab becomes horizontal (in the regime of $z$
where the layering oscillations have died out) only for
$\varepsilon_w\approx 0.65$, however. Thus, for
$\varepsilon_w=0.55$ to $\varepsilon_w=0.61$, where we
approximately have $\theta=90^o$, there is a clear enhancement of
the density in the first two layers adjacent to the walls. Of
course, it would be desirable to study these behaviors varying
also $L_z$ over a wide range, but this has not been attempted
since it would involve a major computational effort. We emphasize
again, that our simulations for these rather thin slit pores are
not suitable to precisely estimate where drying and wetting
transitions occur \{which would show up as macroscopically thick
vapor layers at the walls for the liquid profile $\rho (x_1,z)$ or
liquid layers at the walls for the vapor profile $\rho (x_2,z)$,
respectively\}.

Another caveat that must be made with respect to the quantitative
analysis of our results concern finite size effects associated
with the linear dimension $L_x$. Comparing e.g.~the curves
$\rho(x_1,z)$ for $\varepsilon_w =0.65$ for $L_x=60$ and $L_x=90$ (cf. Fig.~\ref{fig3})
we see that for $L_x=90$ the density clearly is smaller: since the
total particle number was N=9000 in both cases, more particles
were required for $L_x=90$ to create a vapor phase with the proper
density in the box volume, and the remaining particles were only
sufficient for a liquid slab that was too thin to reach the bulk
liquid density in its center. This fact is emphasized in
Fig.~\ref{fig6}, where the density profiles
$\bar{\rho}(x)=\int\limits_1^{L_z-1} \, \rho(x,z)dz/(L_z-2)$ are
shown for 6 values of $\varepsilon_w$. While for $L_x=60$ there is
an (albeit small) region of $x$ where $\bar{\rho}(x)$ is flat in
the center of the liquid slab, for $L_x=90$ the right and left
interfaces clearly are no longer well separated from each other.
This effect becomes more dramatic, of course, if one reduces the
total number of particles in the slit pore (cf. the curve for
$N=4500$, $L_x=45$ and $\varepsilon_w=0.6$, as an example).

Note that the equilibrium conditions for phase coexistence 
under confinement are equal temperature and equal chemical potential 
throughout our systems; the local pressure (and its change due to the 
curvature of the interfaces in Figs. ~\ref{fig1} and ~\ref{fig2}) does 
not play any role in characterizing the equilibrium conditions here. 

We conclude this section by emphasizing that the purpose of our
paper is not the study of wetting and drying transitions, but the
study of liquid-vapor coexistence (and evaporation phenomena) in
slit pores that have nanoscopically small linear dimensions. But
the purpose of the present section was to clarify the equilibrium
properties of phase coexistence under such conditions, and to give
some hints about the finite size effects that one needs to
understand for a proper interpretation of the observed behavior.

\section{Effect of liquid-vapor coexistence in slit pores on the
transport behavior}

Choosing a system in a cubic box geometry of size $L\times L
\times L$ and periodic boundary conditions, it is straightforward
to obtain the self-diffusion constants of the particles from their
mean-square displacement as a function of time, using the Einstein
relation \cite{48,49}. This diffusion constant in the bulk (b)
hence becomes

\begin{equation} \label{eq2}
D_b=\lim\limits_{t \rightarrow \infty} [\langle
[\vec{r}_i(t)-\vec{r}_i (0)]^2 \rangle /(6t) \, .
\end{equation}

Here is understood that the average $\langle \cdots \rangle$
includes an average over all particles (labeled by index $i$ in
Eq.~(\ref{eq2})) in the system, as well as an average over the
origin of time (there exists time translation invariance in
thermal equilibrium, of course). Using Eq.~(\ref{eq2}), the
diffusion constants of bulk vapor and liquid have been obtained,
at vapor-liquid coexistence

\begin{equation} \label{eq3}
D_{\upsilon,b} =1.059 \quad \quad , \quad D_{\ell,b}=0.173 \, \, .
\end{equation}

When we consider a fluid confined in a slit pore, one needs to
consider the following effects: first of all, the mean-square
displacement can diverge only in the $x$- and $y$-directions, but
not in the $z$-direction perpendicular to the walls. 
This type of anisotropy of diffusion in equilibrium was also studied by 
Bock et al. \cite{a}  for a system with patterned walls.
For every
thick slit pores in pure vapor and liquid phases, where the
relative effect of the walls on the diffusion constant can be
neglected, we expect that the diffusion constants in the slit
$(s)$ pore become

\begin{equation} \label{eq4}
D_{\upsilon,s} =(2/3) D_{\upsilon,b} \quad, \quad
D_{\ell,s}=(2/3)D_{\ell, \upsilon} \quad .
\end{equation}

However, the finite linear dimension of the slit pore in the
$z$-direction does introduce slow transients in the behavior of
the mean-square displacement: a particle starting in the center of
the slit pore can also diffuse into the $z$-direction over a
distance $L_z/2$ before the confinement becomes effective. In
fact, such a particle hence can diffuse like in a
three-dimensional bulk system over a time $t_D=(L_z/2)^2 /(6 D_b)$
before the quasi-two-dimensional diffusion sets in. For this
reason, we have defined in terms of the cartesian coordinates
$x_i(t)$, $y_i(t)$ and $z_i(t)$ some effective time-dependence
diffusion constants as follows

\begin{equation} \label{eq5}
D(t)=\langle [x_i(t)-x_i(0)]^2 + [y_i (t) - y_i(0)]^2 +
[z_i(t)-z_i(0)]^2 \rangle / (6t) \quad ,
\end{equation}

\begin{equation} \label{eq6}
D^{(xy)} (t)= \langle [x_i (t) - x_i(0)]^2 + [y_i (t) -y_i (0) ]^2
\rangle /(4t) \quad ,
\end{equation}

\begin{equation} \label{eq7}
D^{(xz)} (t)=\langle [x_i (t) - x_i (0)]^2 + [z_i (t) - z_i (0)]^2
\rangle /(4 t) \quad,
\end{equation}

and

\begin{equation} \label{eq8}
D^{(yz)} (t)= \langle [y_i(t)-y_i(0)]^2 + [z_i (t) - z_i (0) ]^2
\rangle /(4t) \quad.
\end{equation}

As long as there do not occur any interfaces, the symmetry of the
problem requires $D^{(xz)}(t)=D^{(yz)} (t)$, of course, and this
symmetry is in fact nicely obeyed by the numerical data
(Fig.~\ref{fig7}). On the other hand, for $t \ll t_D$ confinement
is not effective, and then the time variation of $D^{(xy)}(t)$,
$D^{(xz)} (t)$ and $D^{(yz)}(t)$ is similar. However, at late
times $D^{(xy)} (t)$ converges to $D_s$ while $D^{(xz)} (t)$ and
$D^{(yz)}(t)$ converge to $D_s/2$, since at late times these mean-
square displacements sample diffusion only in one direction. This
leads to a maximum of $D^{(xz)} (t)$ and $D^{(yz)}(t)$ at
intermediate times (Fig.~\ref{fig7}). Also $D(t)$ for the confined
system exhibits a maximum at about the same time, while the
time-dependence of $D^{(xy)} (t)$ is always monotonic.

From Fig.~\ref{fig7} it is obvious that even in the bulk we must
run the system over a time $\tau \geq 10^2$ to reach saturation at
the asymptotic values of $D$. The diffusion constant in the vapor
is about 6 times larger than in the liquid as noted in
Eq.~(\ref{eq3}), but the times for the mean-square displacements
to converge to these values are about the same. In the bulk, no
cartesian coordinate is distinguished, and hence $D(t)=D^{(xy)}
(t)=D^{(xz)} (t)= D^{(yz)} (t)$. This symmetry properly indeed is
rather nicely fulfilled, and this is only a test of the very good
statistical accuracy of our data.

While for the confined liquid $D^{(xy)} (t)$ follows rather
closely the behavior of the bulk $D(t)$, for $\varepsilon_w=0.59$,
for the confined vapor $D^{(xy)}(t)$ is significantly smaller than
$D(t)$. Presumably, this is due to the fact that the vapor phase
at $\varepsilon_w=0.59$ clearly is rather inhomogeneous (see
Fig.~\ref{fig5}).

The diffusion constants $D(t)$ of the confined system settle down
at $\frac{2}{3} D^{(xy)} (t \rightarrow \infty)$ while the
diffusion constants $D^{(xz)}(t)=D^{(yz)} (t)$ settle down at
$\frac{1}{2} D^{(xy)} (t \rightarrow \infty)$.  These ratios
$2/3$, $1/2$ trivially follow from the normalization of the mean-
square displacements in Eqs.~(\ref{eq5})-~(\ref{eq8}) and the fact
that only mean square displacements of $x$ and $y$ coordinates
diverge. Since the relaxation time $\tau_D$, defined above, which
measures how long it takes for the particles to feel the
confinement does scale inversely with the diffusion constant, it
is plausible that it takes about 6 times longer in the liquid to
reach the asymptotic values of $D(t)$ and $D^{(xy)}(t)$ than it
does in the vapor phase. However, the absolute magnitude of these
times is much larger than expected: using $L_z/2=8$ we would
estimate that in the liquid $t_D$ is of the order of 100 only!

\section{Simulation of Evaporation Processes}

\subsection{Evaporation of Liquid into Vacuum}

 In this section, we consider nonequilibrium relaxation
processes where due to some change of external conditions the size
of a liquid bridge shrinks. E.g., the lateral linear dimension
available for the fluid suddenly increases (we shall not discuss
how such a process could be physically realized in an experiment).
Other conceivable changes of external parameters could involve
changes of temperature, or of the wall-particle interaction, etc.

The first process that is studied is the evaporation of liquid
into ``vacuum'', i.e.~we conceive the situation that a pore is
completely filled with liquid, and due to a sudden change of some
external conditions (e.g.~a confining wall limiting the lateral
extent of the pore in the $x$-direction is removed) additional
pore volume becomes available. The first situation for which this
process is considered refers to pore walls with a purely repulsive
wall-particle interaction (see Sec. 2). In this case we chose an
initial box with linear dimensions $L_x=30$, $L_y=15$, $L_z=15$, walls
being placed at $z=0.5$ and $z=L_z-0.5$, respectively, and
periodic boundary conditions are applied in $x$ and $y$
directions. This system then contains $N=4500$ particles so that
the average density of the liquid in the simulation box is $0.388$
while the density in the center of the box (for $z\approx L_z/2)$
is about 0.598. Then at time $t=0$ the periodic boundary condition
in the $x$-direction is removed, and the system is placed into the
center of a box that is twice as large in $x$-direction, $L_x=60$,
leaving all other linear dimensions and boundary conditions
invariant. For times $t > 0$ a periodic boundary condition in
$x$-direction appropriate for $L_x=60$ is reintroduced. The system
(which was in equilibrium and translationally invariant in
$x$-direction for times $t<0$) now is far out of equilibrium,
because there occurs ``vacuum'' (no fluid particles) for
$0< x <15$, $45< x < 60$, while we have a fluid (inhomogeneous in
$z$-direction because of the repulsive walls, of course) in the
region from $15 < x < 45$ while at $x=15$ and $x=45$ there occur
sharp fluid-vacuum interfaces at $t=0$. Figs.~\ref{fig8}a-d show
the resulting time evolution of the local density $\rho (x,
\bar{z}, t)$ where we have introduced layers 1,2,3,4,5 such that
in layer 1 $\rho(x,z,t)$ is averaged from $z=0$ to $z=1.5$ as well
as from $z=15$ to $z=13.5$, in layer 2 from $z=1.5$ to $z=3$ and
from $z=13.5$ to $z=12$, in layer 3 from $z=3$ to $z=4.5$ and from
$z=12$ to $z=10.5$, in layer 4 from $z=4.5$ to  $z=6$ and from
$z=10.5$ to $z=9$, while layer 5 comprises the central part of the
slit pore (from $z=6$ to $z=9$). Due to the strongly repulsive
wall-fluid particle interaction, almost never any particles occur
in layer 1, and hence only layers 2,3,4 and 5 are shown in
Fig.~\ref{fig8}. One can see that the interfacial profile in
$x$-direction rapidly smoothens, and the density in the central
region of the liquid slab also decreases as fluid particles
evaporate and diffuse into the region of the vapor phase. It
only takes a few hundred MD steps to establish full liquid-vapor
equilibrium (with a homogeneous density of the vapor in
$x$-direction, away from the liquid-vapor interfacial regions) as
the comparison between curves for $t=500$ and $t=20 000$ MD time
steps shows. Note that the fluid particles that ``populate'' the
volume region which in equilibrium form the vapor phase have to
come from the interior of the liquid slab and move through the
region where the liquid-vapor interface forms. The thickness of
this interface gradually increases with time until the equilibrium
interfacial thickness is reached and thus there occur values of
$x$ where the time evolution of the local density is nonmonotonic.
Even for the density fully averaged in $z$-direction, $\rho(x,t) =
L^{-1}_z \int\limits_0^{L_z} \rho (x,z,t)dz$, a clear
non-monotonic time evolution occurs for $x=13.5$ and $x=14.5.$
(Fig.~\ref{fig9}). This happens because as the thickness of the
liquid slab shrinks the interfacial profile moves inward, away
from its initial position at $x=15$, of course. Outside of the
wings of the interfacial profile, e.g.~for $x=10.5$, there is a
monotonous density increase while in the interior of the liquid
slab there is a continuous density decrease.

Of course, one can also study how the density profiles in
$z$-direction changes, at different location in $x$-direction
along the pore (Fig.~\ref{fig10}). One can see here that in the
region where the vapor forms (Fig.~\ref{fig10}a) there is a
monotonic increase of density for all $z$, but the central region
takes longest to equilibrate (for $t=200$ MD time units there is
still a clear deviation from equilibrium). In the interfacial
region, there is a pronounced overshoot of the density in the
center of the pore, Fig.~\ref{fig10}b, while inside the region
when the liquid initially was situated there is a monotonic
density decrease. This is large effect near the interfaces
(Fig.~\ref{fig10}c) and only a small effect in the central region
of the liquid slab (Fig.~\ref{fig10}d).

It is of similar interest to study this process when we
introduce a non-zero fluid-wall interaction $\varepsilon_w$ (Fig.~\ref{fig11}), as
has been studied (with respect to the thermal equilibrium aspects)
already in Sec. 3. It is
interesting to note that there is little effect of $\varepsilon_w$
on the time evolution of the total density $\rho(x,t)$ averaged
over all distances $z$ (Fig.~\ref{fig13}), as long as the final
equilibrium state still contains a thick liquid slab in the center
of the film. This is the case still for $\varepsilon_w=0.59$, but
no longer for $\varepsilon_w=0.9$ (cf. also Sec. 3). Also in the
case of nonzero $\varepsilon_w$ a time of about $\tau=500$ MD time
units suffices to establish the new liquid-vapor equilibrium, with
more or less pronounced precursors of wetting layers at the walls
of the slit pore, as described in Sec. 3. In view of our estimate
of diffusion time scales in Sec. 4, this relatively fast
establishment of equilibrium is perfectly reasonable.

One may also ask the question whether the time scale of
equilibration depends on the $z$-coordinate across the film.
Figs.~\ref{fig10},~\ref{fig11} indicate that such a dependence, if
it exists, is very weak. One could expect, however, that such an
effect should occur for much thicker slit pores under conditions,
where the vapor density is lower in the center of the slit, while
the precursors of the wetting layers then can be several particle
diameters thick. In this situation, the relaxation time in the
liquid layers adjacent to the walls could be much larger than in
the vapor region. In our case, however, even for
$\varepsilon_w=0.9$ where pronounced layering in the dense regions
that build up close to the walls is observed (Fig.~\ref{fig11}) no
significant slowing down has been detected.

\subsection{Evaporation of Vapor into Vacuum}

In the previous subsection we have considered the situation that
for a slit pore completely filled with liquid an additional volume
becomes available, into which evaporation can take place, and have
presented data that illustrate how liquid-vapor interfaces form
and a vapor-liquid phase equilibrium is established. In the
present subsection, we consider the alternative scenario of a slit
pore, in which such a vapor-liquid phase equilibrium already
occurs, and by an external operation additional volume for the
vapor phase becomes available. Of course, when the vapor spreads
into the part of the volume which initially is empty, the average
vapor density decreases, and the vapor no longer is in equilibrium
with the liquid phase, with which it has coexisted in equilibrium
for time $t < 0$. As a result, also liquid will evaporate again,
driven by the density gradient that occurs in the vapor region,
until the density in the whole region taken by the vapor has
adjusted to the value the vapor density must have at coexistence
with the liquid in thermal equilibrium at the chosen temperature
and boundary conditions (i.e., value of $\varepsilon_w$) at the
walls of the confining slit pore.

Using the final equilibrium states for the system with linear
dimensions $L_x=60$, $L_y=15$, $L_z=15$ with purely repulsive
walls as an initial condition, we have increased the linear
dimension in $x$-direction from $L_x=60$ to $L_x=72$. Due to this
rather modest increase of $L_x$, the liquid slab in the center of
the system does not evaporate completely, but simply gets only a
bit smaller, as a consideration of the average profile $\rho(x,t)$
shows (Fig.~\ref{fig14}). One sees that the strong density
gradient (where originally the average density $\rho(x,t=0)$ jumps
from about 0.092 to zero ) at $x=6$ and $x=66$ (the vacuum takes
the region $0 \leq x < 6$ and $66 < x \leq 72$ at $t=0$) smoothes
out already during first 10 MD time units, and for $t=100$ the
vapor density in the region $0 \leq x < 12$ and $60 < x \leq 72$
is almost independent of $x$, but the density in this region is
distinctly smaller than the critical coexistence density. The
thickness of the liquid slab has remained almost unchanged. Thus,
in this initial regime of times it is basically the vapor present
in the original system that has spread out into the empty region,
which is understandable since the diffusion constant in the liquid
region is much smaller, and also the driving force for evaporation
of the liquid slab is clearly not very large. From
Fig.~\ref{fig14} one can see that the shape of the interfacial
profile (at least in its central part) does not change with time,
it is only the interface position (which we may characterize
precisely from the inflection point of $\rho(x,t)$ in
Fig.~\ref{fig14}) that shifts in the time regime from $t \approx
100$ to $t\approx500$ with approximately constant velocity, while
for $t \geq 500$ the vapor density for $x \leq 12$ and $x \geq 60$
starts to saturate, and then the interface velocity also decreases
to zero.

As in the cases studied so far, we have again tried to obtain more
detailed information on the spatially resolved data. At the
slabs centered at $x=0.5$, 12.5, 24.5 and 34.5 (all slabs have a
width $\Delta x =1.0$) we have followed the time evolution of
$\rho (x, \bar{z}, t)$, defining the positions $\bar{z}$ of the 5
``slices'' in $z$-direction in the same way as in the equilibrium
case. Since 100 runs needed to be followed for 10000 MD time
units, this part of the study involves a major computational
effort, although the statistical fluctuations necessarily are
still rather large (Fig.~\ref{fig15}). One can see that for $t
\geq 2000$ there is no significant relaxation any more, while
times $t \leq 500$ clearly are not enough to fully establish
equilibrium. While in the center of the interfacial region
($x=24.5$) the relaxation is slow but the density there decreases in
a monotonic fashion, a nonmonotonic density relaxation occurs in
the wings of the profile ($x=12.5)$ far away from the walls. In
the regime closer to the walls (layers 1 and 2) the relaxation is
always much faster, however.

\subsection{Transient Diffusion during Evaporation Processes}

It is possible in the simulation to ask the question how the
presence of an evaporation process shows up in the time-dependence
of the mean-square displacement of the particles. Experimentally,
such a question could be asked e.g.~for colloid-polymer mixtures
\cite{65}, where a vapor liquid type phase separation occurs for
suitable conditions, and it is possible to follow the motion of
individual colloid particles with fluorescent labels \cite{66}. In
Fig.~\ref{fig16}, we show representative results for the
time-dependent diffusivities defined in
Eqs.~(\ref{eq5})-~(\ref{eq8}) for an evaporation simulation where
$L_x$ was changed from $L_x=30$ to $L_x=60$ at $t=0$. Unlike the
situation discussed in Sec. 4, where the choice of the origin of
time did not play any role since the system in equilibrium obeys
time translation invariance, this is no longer the case now: all
quantities in Eqs.~(\ref{eq5})-~(\ref{eq8}) depend on two times
now, the time chosen for the origin $t=0$ there, and the time $t$
elapsed since then. We have not attempted to study this
non-stationary transport problem in full detail, however, but
focus only on the case where the deviation from equilibrium is
strongest, i.e.~when the origin of time chosen in
Eqs.~(\ref{eq5})-~(\ref{eq8}) coincides with the time where the
change of $L_x$ is performed.

While the initial ballistic regime (where $D(t) \propto t)$ that
one can recognize for $1 \leq t\leq 6$ in Fig.~\ref{fig16}, is not
much affected by the fact that the system is out of equilibrium,
for $t> 6$ a very different behavior occurs: $D^{(yz)} (t)$
becomes slower for some transient time period, while for $t > 20$
another speed-up occurs, and near $t=100$ a maximum of $D^{(yz)}
(t)$ occurs, followed by a slow decay to the asymptotic value. The
precise height and location of this maximum depend on the choice
of $\varepsilon_w$ slightly. All other time-dependent diffusion
constants do contain also mean-square displacements in
$x$-direction, and they show a strong speed up already at times $t
> 6$. We attribute this speed up to the drift that occurs in
$x$-direction: since it is more likely for the particles to move
in $x$-direction than in any other direction, $D ^{(yz)}(t)$
exhibits a slower increase than in the ballistic regime. This
nonequilibrium enhancement of $D(t)$ and the various choices for
$D^{(\alpha \beta)} (t)$ [$\alpha, \beta=x,y,z$] shown in
Fig.~\ref{fig16} leads to maxima in most of the mean-square
displacements that are more pronounced than the corresponding data
in the confined pure phases in equilibrium (Fig.~\ref{fig7}). The
diffusion constants that the systems relax to, which show up as
the flat plateaus in Fig.~\ref{fig16} for $t \geq 2000$, have
values in between that of confined pure liquid and vapor phases,
consistent with the (qualitative) expectation. Of course, we are
far from a quantitative understanding of transport phenomena
during relaxation in such inhomogeneous two-phase systems in
confined geometry, however.

\section{Conclusions}

Fluids confined in slit-like pores can form liquid bridges coexisting
 with vapor at temperatures below the critical point. In the present work,
 we have studied for a simple Lennard-Jones model of a fluid the static
 structure of such liquid bridges, for several typical cases of fluid-wall interactions, corresponding 
to (incomplete) drying and wetting conditions.
We have used molecular dynamics simulations to explore the partial
 (or complete) evaporation of such bridges resulting from changes of external thermodynamic control 
parameters and we have discussed the interplay of finite size effects (associated 
with the finite width of the slit pore, or with the finite linear dimensions
 of the liquid bridge in the directions parallel to the slit walls, or 
both) and surface effects due to the walls.

\begin{figure}
\begin{center}
\includegraphics[width=8cm,clip,angle=-0]{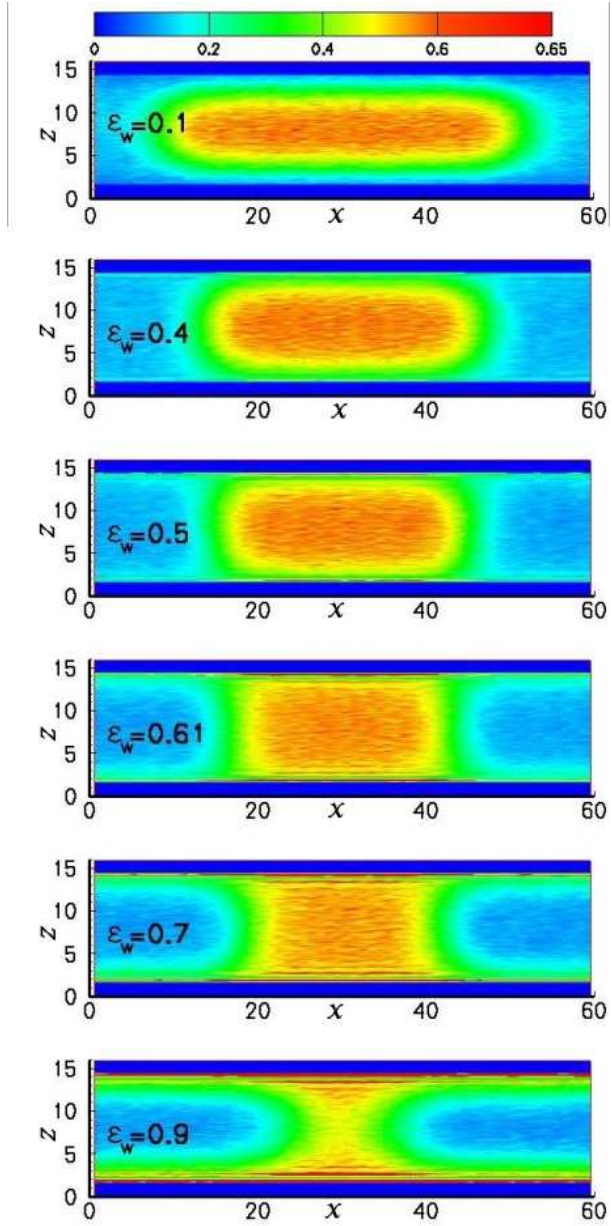}
\caption{(Color) Density distribution $\rho(x,z)$ for systems with linear
dimensions $L_x=60$, $L_y=20$ and $L_z=16$, for 6 values of
$\varepsilon_w= 0.1,0.4,0.5,0.61, 0.7, 0.9$
(from top to bottom). The $z$-axis is oriented along
the ordinate and the $x$-axis along the abscissa. The value of the density $\rho$
corresponds to the color code, as shown by the bar on the right
side of the plots. The temperature is
$T^*=k_BT/\varepsilon=0.9366$. The system was simulated for
20000 time units (cf. description in Sec. 2), and then averages were taken over last 15000 MD time
units $\tau$.\label{fig1}}
\end{center}
\end{figure}

\begin{figure}
\begin{center}
\includegraphics[width=8cm,clip,angle=-0]{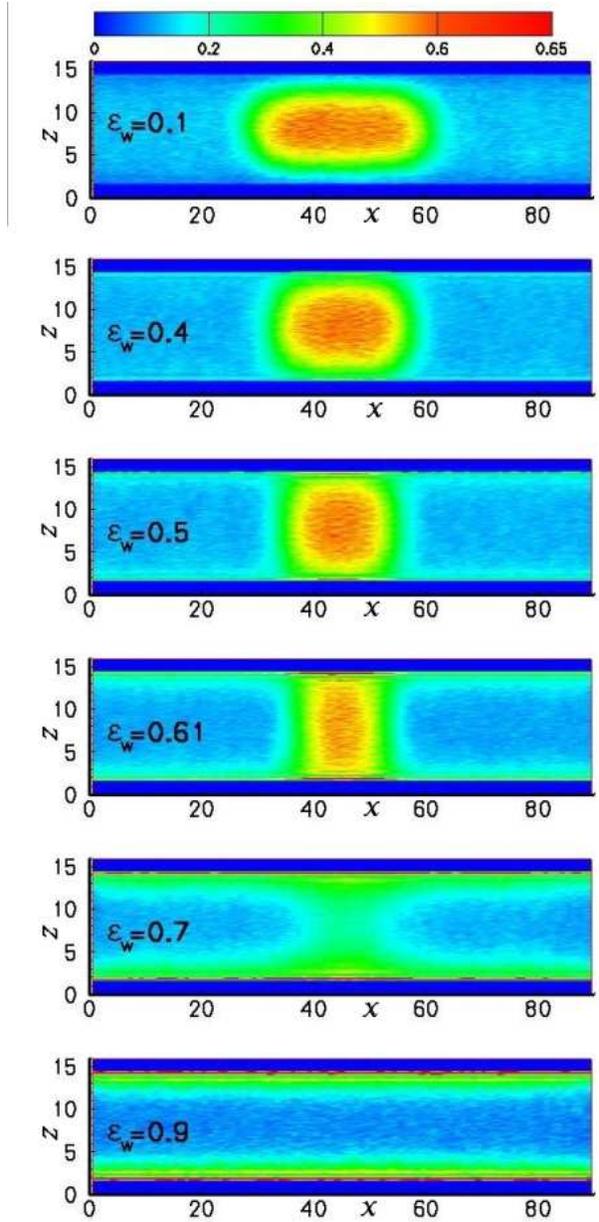}
\caption{(Color) Same as Fig.~\ref{fig1}, but for $L_x=90$ instead of
$L_x=60$.\label{fig2} Note that the scale along
the $x$-direction is compressed relative to the
scale for the $z$-direction.}
\end{center}
\end{figure}

\begin{figure}
\begin{center}
\includegraphics[width=8cm,clip,angle=-0]{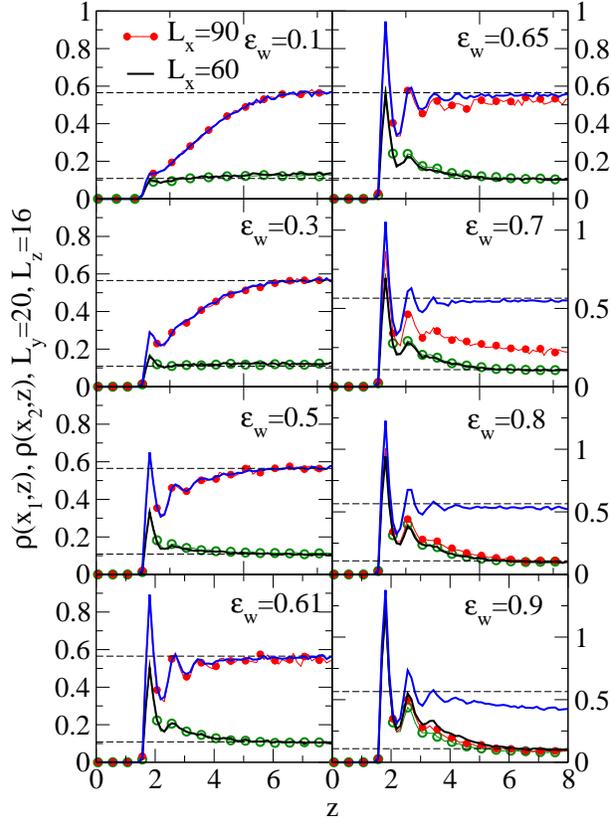}
\caption{(Color online) Density profiles $\rho(x_1,z)$ and $\rho(x_2,z)$ plotted
vs. $z$ for 8 values of
$\varepsilon_w=0.1, 0.3, 0.5, 0.61$ (from top to
bottom, left column) and $\varepsilon_w=0.65, 0.7, 0.8,0.9$ (from
top to bottom, right column). 
Solid lines and lines with circles represent density profiles for the system
 $60 \times 20 \times 16$ (cf. Fig.~\ref{fig1}) 
and $90 \times 20 \times 16$ (cf. Fig.~\ref{fig2}), respectively. 
In each frame of the panel the upper
curve shows $\rho(z,x_1)$ with $x_1$ being defined via
$\rho(x_1,z)=\int\limits_{25}^{35} \rho (x,z) dx /2$, and the
lower curve shows $\rho(x_2,z)$, with $x_2$ being defined via
$\rho(x_2,z)= [\int\limits^3_0 \rho(x,z) dx +
\int\limits^{60}_{57} \rho (x,z) dx]/6$. Thus, the upper curve
shows the density profile along a cut through the center of the
liquid slab, while the lower curve shows the density profile
through the vapor region far away from the vapor-liquid interface.
These averages are carried out for time above $t>5000$ over 15000 MD time units $\tau$. 
Horizontal broken straight lines show bulk  $\rho_l$ and $\rho_g$, respectively.
\label{fig3}}
\end{center}
\end{figure}

\begin{figure}
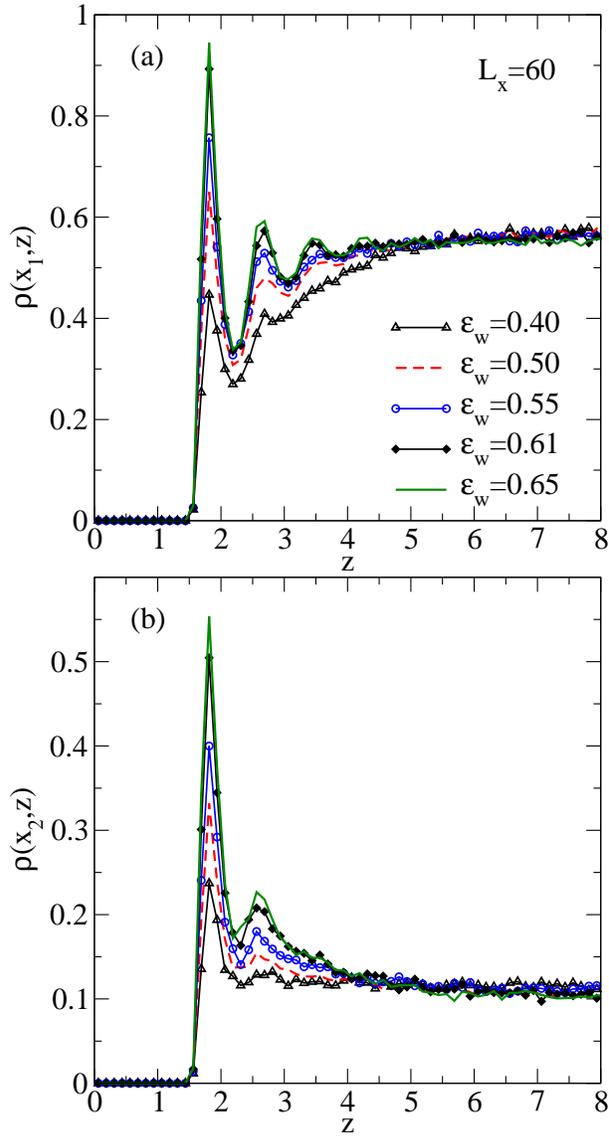

\begin{center}
\includegraphics[width=8cm,clip,angle=-0]{FIG_4a.eps}
\includegraphics[width=8cm,clip,angle=-0]{FIG_4b.eps}
\caption{(Color online) Magnified plot of finely resolved density profiles in liquid 
$\rho(x_1,z)$ (a) and vapor $\rho(x_2,z)$ (b) plotted vs. $z$ for
$\varepsilon_w=0.4,0.5,0.55,0.61$ and $0.65$, as indicated in the
figure, for $L_x=60$, all other parameters
being the same as in Figs.~\ref{fig1}-\ref{fig3}. Note that the
cells for the averaging have a width $\Delta z=0.125$ in
$z$-direction.\label{fig5}}
\end{center}
\end{figure}

\begin{figure}
\begin{center}
\includegraphics[width=8cm,clip,angle=-0]{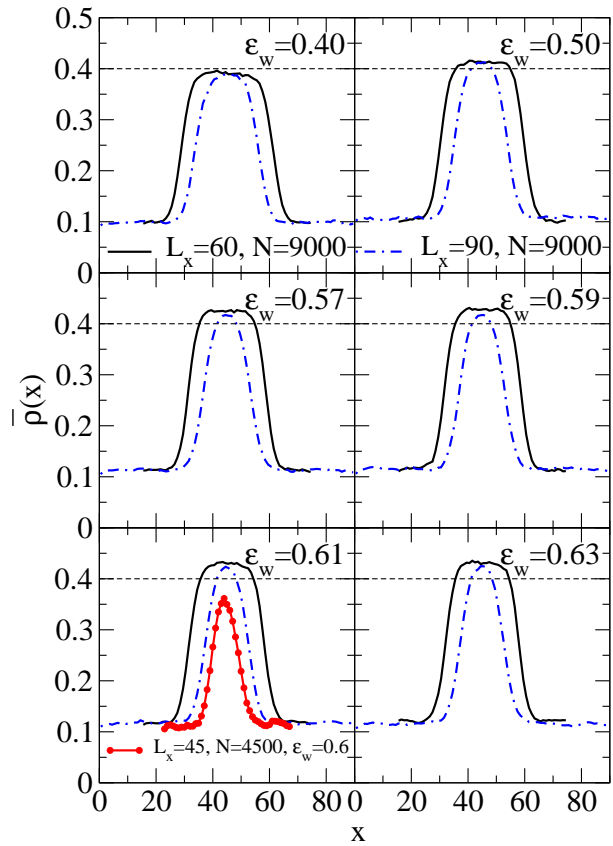}
\caption{(Color online) Density profiles $\bar{\rho}(x)$ averaged in
$z$-direction plotted vs. $x$ for $L_x=60$ and $L_x=90$, for 6
values of $\varepsilon_w$, as indicated \label{fig6}}.
\end{center}
\end{figure}

When one considers the density variation in the direction perpendicular 
 to the confining walls, one finds that both in the liquid phase and in the 
vapor phase the density approaches the values of bulk liquid and vapor 
at the coexistence curve, after about only 5 Lennard-Jones diameters, if one 
stays away from the region where liquid-vapor interfaces run across 
the slit pore (this condition also requires, of course, that the liquid bridge is thick 
enough that the two interfaces separating it from the vapor phase are 
non-interacting). This finding holds for all choices of the wall-fluid interaction (only 
when the slit width D gets several orders of magnitude larger than the Lennard-Jones diameter and 
if one is very close to conditions of complete wetting or complete drying, would be a larger inhomogeneity 
in the direction perpendicular to the wall expected). 
On the other hand, when the two interfaces between the liquid bridge and the vapor are close enough together so that 
their interaction cannot be neglected (e.g. the case $\varepsilon_w=0.9$ in Figs. 
~\ref{fig1}, ~\ref{fig3}; $\varepsilon_w=0.7$ in Figs.
~\ref{fig2}, ~\ref{fig3}), the density inside the liquid bridge remains smaller 
than in the bulk, and one can observe a smooth crossover to the state where 
the liquid "bridge" rather should be described as two wall-attached droplets opposite of each other 
(Fig. ~\ref{fig2}, $\varepsilon_w=0.7$). Changing the thermodynamic conditions,
 in this way a smooth crossover from states containing a bridge to states without a bridge 
(Fig. ~\ref{fig2}, $\varepsilon_w=0.9$) are possible, and no sharp phase transitions 
(in the sense of a singular behavior of thermodynamic potentials or their derivatives) are 
involved, when all linear dimensions considered remain finite.

\begin{figure}
\begin{center}
\includegraphics[width=8cm,clip,angle=-0]{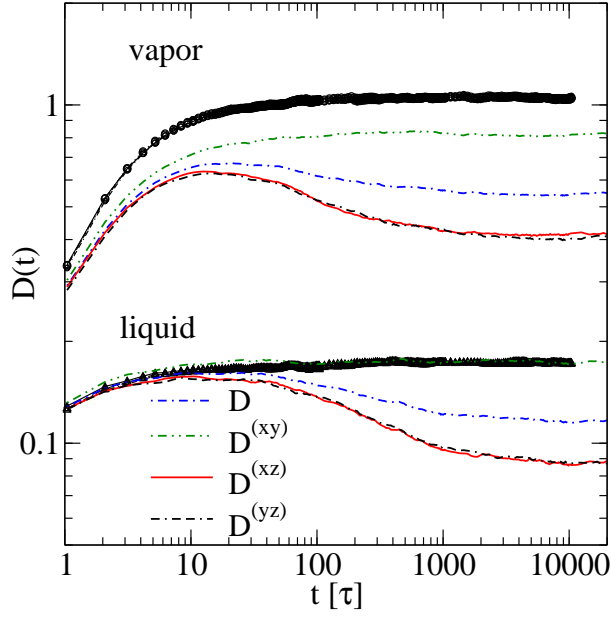}
\caption{(Color online) Log-log plot of time-dependent diffusion constants
$D(t)$, $D^{(xy)} (t)$, $D^{(xz)}(t)$ and $D^{(yz)}(t)$ vs. time,
for systems at $T^*=0.9366$ both in the vapor phase (which in the
bulk has a density $\rho=0.10866$) and in the liquid phase (which in
the bulk has a density $\rho=0.565032$). Both data for bulk systems
(linear dimensions $27 \times 20 \times 16$ for the pure liquid at
coexistence and $140 \times 20 \times 16$ for the pure vapor phase
at coexistence, with periodic boundary conditions in all three
directions) and for confined systems in single-phase states
(confined by walls with $\varepsilon_w=0.59$, choosing linear
dimensions $34\times 20\times 16$ for the liquid and $240\times 20
\times 16$ for the vapor) are included. Lines with circles correspond to
bulk results for vapor, lines with triangles - to bulk results for liquid. 
Lines without symbols correspond to data for confined systems. \label{fig7}}
\end{center}
\end{figure}

\begin{figure}
\begin{center}
\includegraphics[width=8cm,clip,angle=-0]{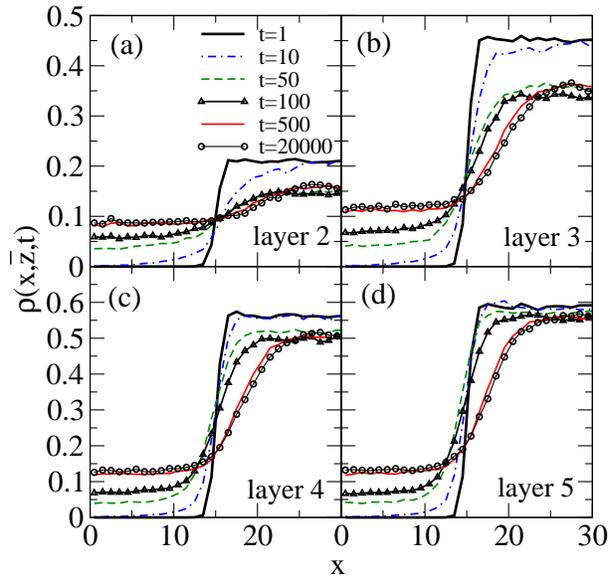}
\caption{(Color online) Time evolution of the average density $\rho(x,\bar{z},t)$
during the evaporation of liquid into vacuum 
after the change of linear dimension $L_x=30$ to $L_x=60$ for
layer 2 (a), 3 (b), 4 (c) and 5 (d). The region of $z$ over which
$\rho(x,z,t)$ is averaged in the different layers is explained in
the main text. Different curves indicate time $t$ after the volume
change, as indicated. These data result from averaging over 100
independent and equivalent runs. \label{fig8}}
\end{center}
\end{figure}

\begin{figure}
\begin{center}
\includegraphics[width=8cm,clip,angle=-0]{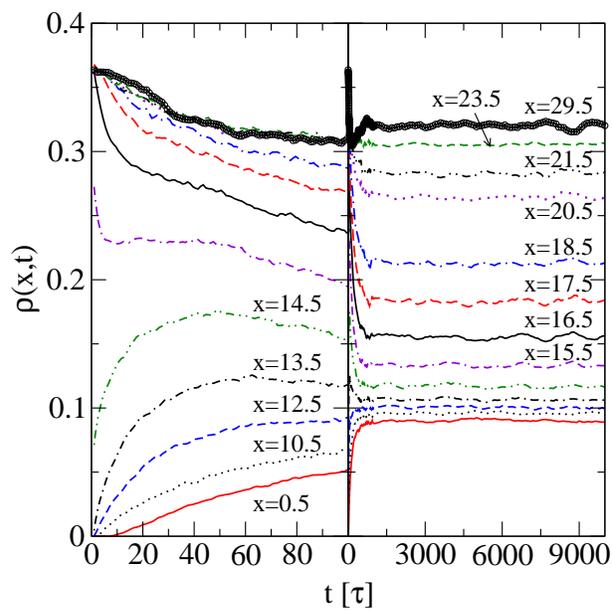}
\caption{(Color online) Evaporation of liquid into vacuum. Time dependence of the density $\rho(x,t)$ averaged in
$z$-direction plotted versus time for short time scales (left
part) and over long times (right part). Different curves show a
few values of $x$, as indicated in the figure. \label{fig9}}
\end{center}
\end{figure}

\begin{figure}
\begin{center}
\includegraphics[width=8cm,clip,angle=-0]{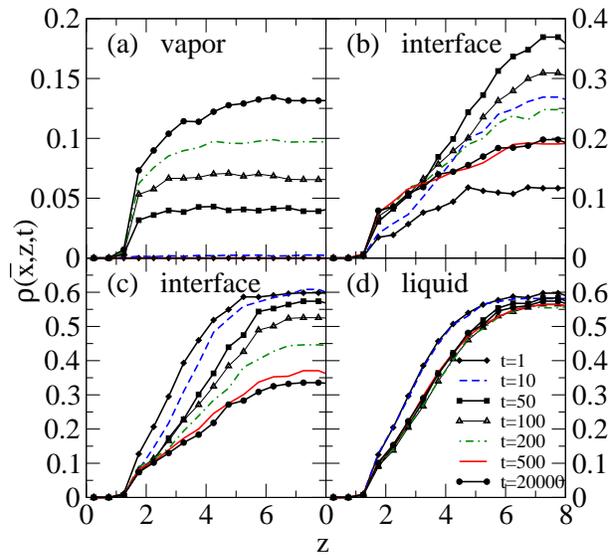}
\caption{(Color online) Profiles $\rho(\bar{x}, z,t)$ plotted vs. $z$ for $x$
averaged from 0 to 3 and 57 to 60 (a), from 14 to 15 and 45 to 46
(b), from 17 to 18 and 47 to 48 (c) and from 27 to 33 (c).
\label{fig10}}
\end{center}
\end{figure}

\begin{figure}
\begin{center}
\includegraphics[width=8cm,clip,angle=-0]{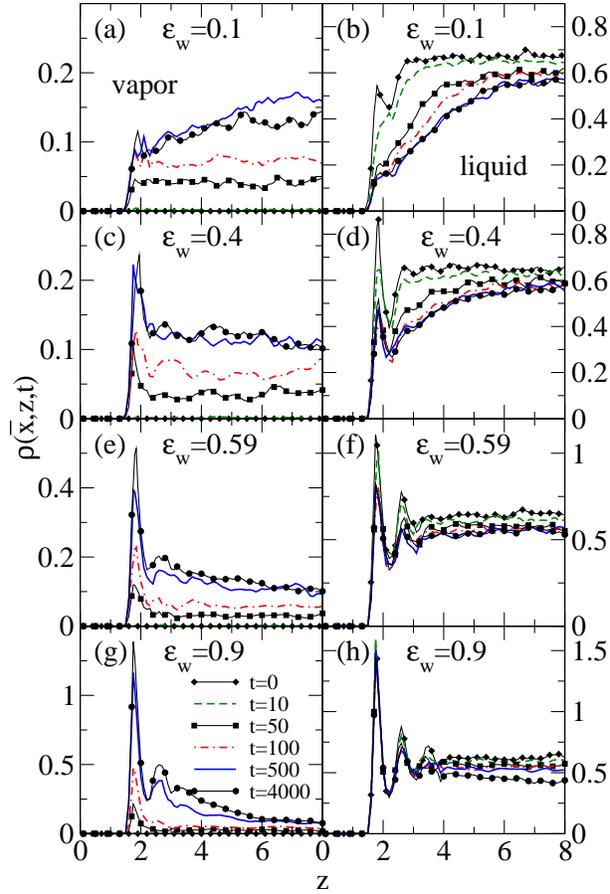}
\caption{(Color online) Profiles $\rho(\bar{x}, z,t)$ plotted vs. $z$ for $x$
averaged from 0 to 5 and 55 to 60(left column) and from 25 to 35 (right column), for
$\varepsilon_w=0.1$ (a,b), $\varepsilon_w=0.4$
(c,d), $\varepsilon_w=0.59$ (e,f) and $\varepsilon_w=0.9$
(g,h) choosing now a box of linear dimensions $60
\times 20 \times 16$, and different times $t$ as indicated, after
a change of $L_x$ from $L_x=30$ to $L_x=60$ was
performed.\label{fig11}}
\end{center}
\end{figure}

\begin{figure}
\begin{center}
\includegraphics[width=8cm,clip,angle=-0]{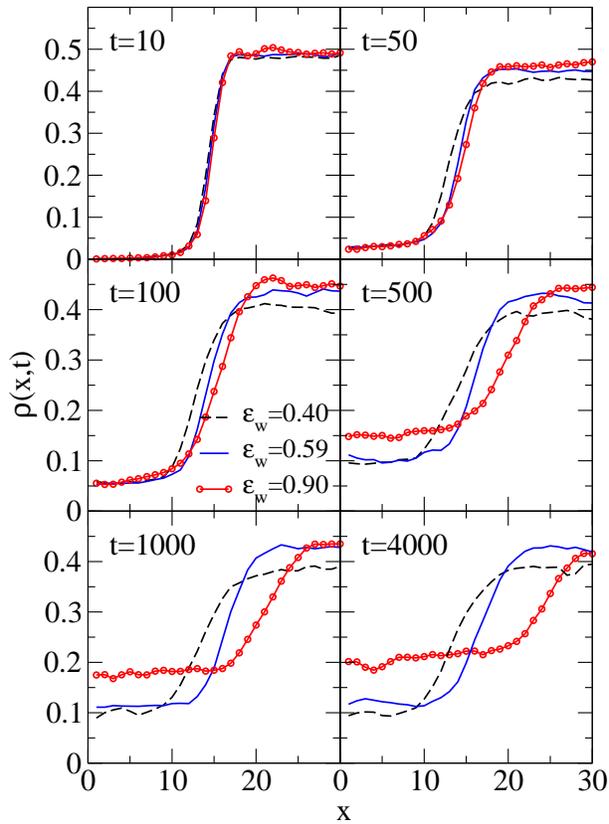}
\caption{(Color online) Total density $\rho(x,t)$ averaged over the $z$-direction
across the slit pore, for a box of linear dimensions $L_x \times
20 \times 16$, at times $t$ after at time $t=0$ a change of $L_x$
from $L_x=30$ to $L_x=60$ was performed. Three choices of
$\varepsilon_w$ are shown: $\varepsilon_w=0.4$, 0.59 and 0.9. \label{fig13}}
\end{center}
\end{figure}

\begin{figure}
\begin{center}
\includegraphics[width=8cm,clip,angle=-0]{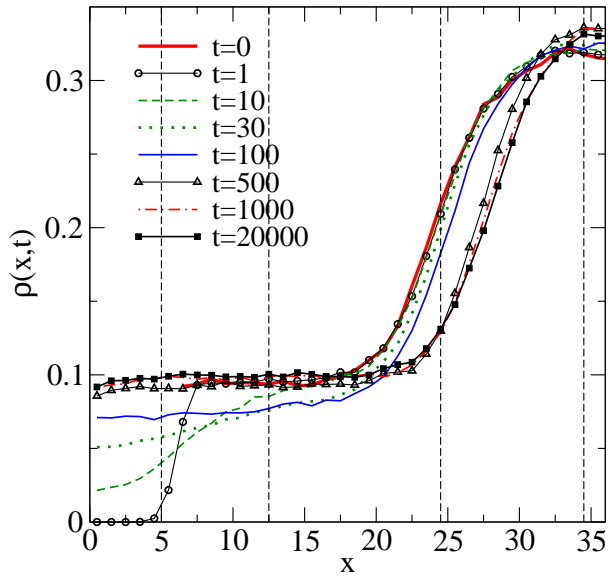}
\caption{(Color online) Time dependence of the density $\rho(x,t)$ averaged in
the $z$-direction across the slit pore, for a system of size $L_x
\times 15 \times 15$, for which at time $t=0$ a change from
$L_x=60$ to $L_x=72$ is performed (evaporation of coexisting liquid and vapor into vacuum).
 The interaction between the
particles forming the walls and the fluid particles is purely
repulsive. Different curves indicate the various times, as
indicated. The vertical straight lines highlight the positions
(see text) where the spatially resolved time dependence is
analyzed in Fig.~\ref{fig15}. \label{fig14}}
\end{center}
\end{figure}

\begin{figure}
\begin{center}
\includegraphics[width=8cm,clip,angle=-0]{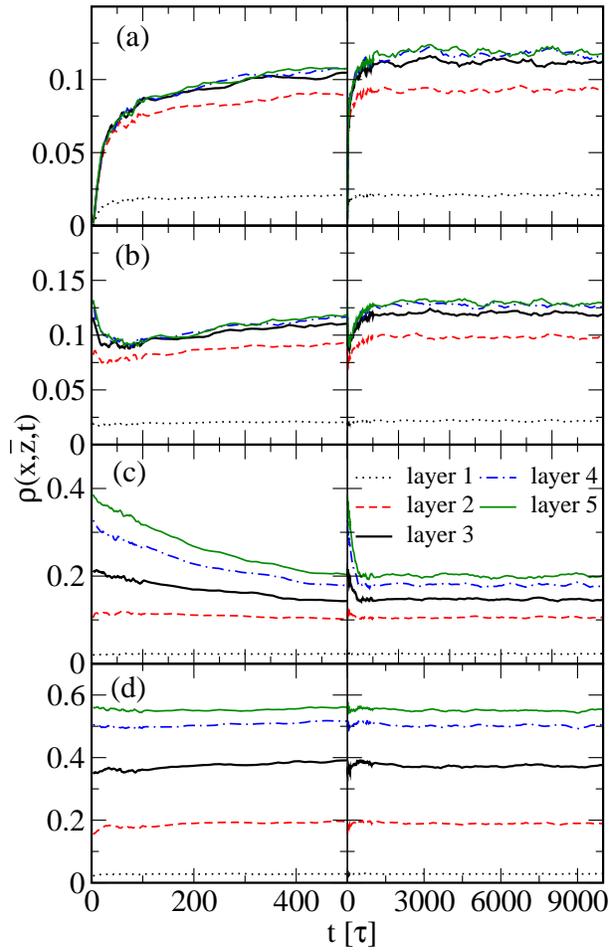}
\caption{(Color online) Plot of $\rho(x, \bar{z}, t)$ vs. $t$ for $x=0.5$ (a),
12.5(b), 24.5 (c) and 34.5 (d). Each panel shows five values of
$\bar{z}$, as indicated in part (c). The left part shows always
the initial relaxation  ($0 \leq t \leq 500$), the right part
shows the full interval of times ($0 \leq t \leq 10000)$ over
which the simulations were extended. For further explanations see
text. \label{fig15}}
\end{center}
\end{figure}

\begin{figure}
\begin{center}
\includegraphics[width=8cm,clip,angle=-0]{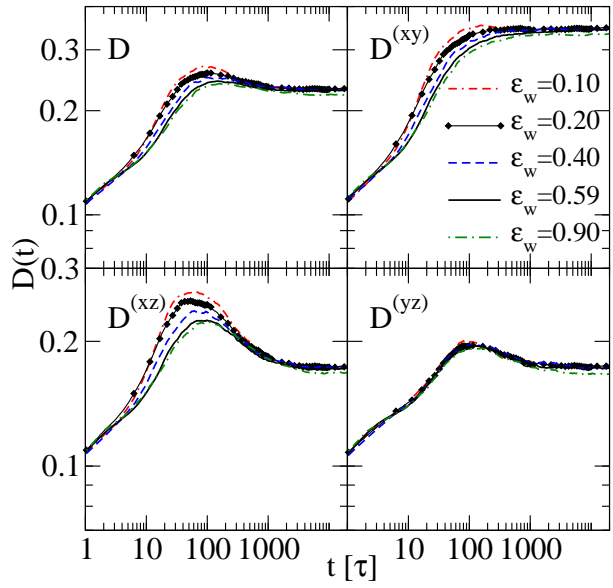}
\caption{(Color online) Effective time-dependent diffusion constants $D(t)$, $D
^{(xy)} (t)$, $D^{(xz)} (t)$ and $D^{(yz)} (t)$ plotted vs. time
on a log-log plot, for several choices of $\varepsilon_w$, as
indicated. The origin of time $t=0$ used in the definitions,
Eqs.~(\ref{eq5})-~(\ref{eq8}), was chosen coincident with the
change from $L_x=30$ to $L_x=60$, where evaporation of the
confined liquid into vacuum starts. \label{fig16}}
\end{center}
\end{figure}

For the conditions studied, there is a significant (but not too strong) dynamic asymmetry 
between the coexisting phases in the bulk (the diffusion constant of the vapor is about 
6 times larger than that of the liquid, Fig. ~\ref{fig7}). When one considers either pure 
vapor or pure liquid phases in confinement, the crossover from three-dimensional to 
quasi-two-dimensional diffusion already causes slow transients in the mean square displacements, 
Eq.~(\ref{eq6}) of the order of several thousand MD time units (physically, this may correspond to 
about 10 nanoseconds). It turns out that this time-scale associated with diffusion 
in confined geometry in equilibrium is larger than the time scale on which 
evaporation processes take place (Figs. ~\ref{fig10}-~\ref{fig15}).
Both the evaporation of liquid into vacuum and of vapor into vacuum 
is essentially completed after a few hundred MD time units already. 
When the vapor that evaporates into the vacuum still coexists with a 
liquid bridge in the center of the slit pore, the liquid bridge must somewhat 
shrink to maintain local thermal equilibrium: in this way 
the establishment of a density gradient in the vapor region of the system can be avoided. 
During the evaporation process, some of the mean-square displacements show super-diffusive 
behavior (Fig. ~\ref{fig16}). 

Of course, our observations are only a first step towards the full clarification of the problem: 
it would be interesting to vary both parallel and perpendicular linear dimensions over at least a decade systematically, to clarify under which conditions the evaporation process becomes much slower (which then would 
be relevant for applications). Also a study of the variation with temperature would be interesting. 
However, all such extensions need substantial computer resources, and must be left to future work. 
However, we hope that the present work stimulates both the development of 
phenomenological analytical work and experimental studies on these issues.

{\bf Acknowledgements}: The first author (K.B.) acknowledges financial support from 
 the Alexander von Humboldt Foundation.

\clearpage

\clearpage

\clearpage

\clearpage
%\begin{figure}
%\caption{Same as Fig.~\ref{fig3}, but for $L_x=90$ instead of
%$L_x=60$. \label{fig4}}
%\end{figure}

\clearpage

\clearpage

\clearpage

\clearpage

\clearpage

\clearpage

%\begin{figure}
%\begin{center}
%\includegraphics[width=12cm,clip,angle=-0]{FIG_12.eps}
%\caption{Same as Fig.~\ref{fig11}, but for $\varepsilon_w=0.4$
%(a,b), $\varepsilon_w=0.59$ (c,d) and $\varepsilon_w=0.9$
%(e,f).\label{fig12}}
%\end{center}
%\end{figure}

\clearpage

\clearpage

\clearpage

\clearpage

\end{document}